\documentclass[12pt]{iopart}
\newcommand{\be}{\begin{equation}}
\newcommand{\ee}{\end{equation}}
\newcommand{\ben}{\begin{eqnarray}}
\newcommand{\een}{\end{eqnarray}}

\newcommand{\anf}{``}

\begin{document}
\title{Chaplygin gas with non-adiabatic pressure perturbations}

\author{Winfried Zimdahl\dag \footnote{Electronic address:
zimdahl@thp.uni-koeln.de}\ and Julio C. Fabris\ddag
\footnote{Electronic address: fabris@cce.ufes.br}  }

\address{\dag\ Institut f\"ur Theoretische Physik, Universit\"at zu
K\"oln, D-50937 K\"oln, Germany}

\address{\ddag\ Departamento de F\'{\i}sica
Universidade Federal do Esp\'{\i}rito Santo, \\CEP29060-900
Vit\'oria, Esp\'{\i}rito Santo, Brazil}


%

\begin{abstract}
Perturbations in a Chaplygin gas, characterized by an equation of
state $p = -A/\rho$, may acquire non-adiabatic contributions if
spatial variations of the parameter $A$ are admitted. This feature
is shown to be related to a specific internal structure of the
Chaplygin gas. We investigate how perturbations of this type
modify the adiabatic sound speed and influence the time dependence
of the gravitational potential which gives rise to the Integrated
Sachs-Wolfe effect in the anisotropy spectrum of the cosmic
microwave background.
\end{abstract}

\maketitle

\section{Introduction}

Since 1998 \cite{SN} a growing number of observational data has
backed up the conclusion that the expansion rate of our present
Universe is increasing. According to our current understanding on
the basis of Einstein's General Relativity, such a dynamics
requires a cosmic substratum with an effective negative pressure.
To clarify the physical nature of this substratum is one of the
major challenges in cosmology. Most of the approaches in the field
rely on a two-component picture of the cosmic medium with the (at
present) dynamically dominating \anf Dark Energy" (DE), equipped
with a negative pressure, which contributes roughly 70\% to the
total energy density, and with pressureless (Cold) \anf Dark
Matter" (CDM), which contributes roughly 30\% (see, e.g.
\cite{Dark} and references therein). Usually, these components are
assumed to evolve independently, but more general interacting
models have been considered as well \cite{IQ}. A single-component
model which has attracted some interest as an alternative
description is a Chaplygin gas \cite{Chaplygin}. The Chaplygin gas
which is theoretically based in higher dimensional theories
\cite{Jackiw}, has been considered as a candidate for a unified
description of dark energy and dark matter
\cite{Kamen,Julio,bento1,Finelli1,bento2}. Its energy density
smoothly changes from that of matter at early times to an almost
constant value at late times. It interpolates between a phase of
decelerated expansion, necessary for structure formation to occur,
and a subsequent period in which the dynamically dominating
substratum acts similarly as a cosmological constant, giving rise
to accelerated expansion. The Chaplygin gas combines a negative
pressure with a positive sound velocity. This sound speed is
negligible at early times and approaches the speed of light in the
late-time limit. A sound speed of the order of the speed of light
has implications which apparently disfavor the Chaplygin gas as a
useful model of the cosmic medium. In particular, it should be
connected with oscillations of the medium on small (sub-horizon)
scales. The fact that the latter are not observed has led the
authors of \cite{Sandvik} to the conclusion that Chaplygin gas
models of the cosmic medium are ruled out  as competitive
candidates. Similar results were obtained from the analysis of the
anisotropy spectrum of the cosmic microwave background
\cite{Finelli2,Bean}, except possibly for low values of the Hubble
parameter \cite{bento3}. However, these conclusions rely on the
assumption of an adiabatic cosmic medium. It has been argued that
there might exist entropy perturbations, so far not taken into
account, which may change the result of the adiabatic perturbation
analysis \cite{NJP,Reis}. A problem here is the origin of
non-adiabatic perturbations which should reflect the internal
structure of the cosmic medium. The latter is unknown but it may
well be more complicated then suggested by the usually applied
simple (adiabatic) equations of state. Generally, non-adiabatic
perturbations will modify the adiabatic sound speed. The speed of
sound has attracted interest as a tool to discriminate between
different dark energy models \cite{Erickson,DeDeo,Bean2}. It has
been argued that (anti-)correlations between curvature and
isocurvature fluctuations may provide a mechanism to modify the
(adiabatic) sound speed of the dark energy \cite{Moroi,Gordon}. A
reduced sound speed (compared with the velocity of light) in
quintessence models may lead to a suppression of the lowest
multipole moments of the CMB anisotropy spectrum (which are
usually enhanced by the Integrated Sachs-Wolfe (ISW) effect
\cite{DeDeo,Finelli3,Gordon}). The point here is, that a change of
the sound speed will affect the time dependence of the
gravitational potential on large perturbation scales. It is this
time-dependence that gives rise to the ISW effect \cite{SW}. A
detection of the ISW, representing an independent confirmation of
the existence of dark energy, has recently been reported by
\cite{Nolta,Boughn,Vielva}. Since the CMB data suggest reduced
power of the lowest multipoles (compared with the prediction of
the $\Lambda$CDM model), different reduction mechanisms have been
discussed in the literature (see, e.g. \cite{Contaldi}), among
them also approaches that require \anf new physics" (see, e.g.,
\cite{Bastero,Tsujikawa}), which will not be considered, however,
in this paper.

The purpose of this paper is to study a simple model of
non-adiabatic perturbations in a Chaplygin gas. Starting with a
two-component description of the cosmic medium we demonstrate that
any Chaplygin gas in a homogeneous and isotropic Universe can be
regarded as being composed of another Chaplygin gas which is in
interaction with a pressureless fluid such that that the
background ratio of the energy densities of both components is
constant. Spatial perturbations of this ratio then give rise to
non-adiabatic pressure perturbations which may be related to
fluctuations of the equation of state parameter $A$. We
investigate the influence of these perturbations on the sound
speed and on the time dependence of the gravitational potential.

The paper is organized as follows. In Section \ref{two-fluid
dynamics} we introduce the necessary definitions and relations to
describe the dynamics of an interacting two-fluid model both in a
homogenous and isotropic background and on the perturbative level.
Section \ref{Chaplygin} starts by recalling basic relations for
the Chaplygin gas. Then we show that any one-component Chaplygin
gas in a homogeneous and isotropic universe is equivalent to an
interacting two-component system, where one of the components is
another (different parameters) Chaplygin gas and the second
component is dust. The interaction is such that the ratio of the
energy densities of both components remains constant in time. On
the basis of this two-component interpretation we introduce
non-adiabatic pressure perturbations in section
\ref{non-adiabatic} by allowing the ratio of the energy densities
of both components to fluctuate. Relating these fluctuations of
the density ratio to fluctuations of the equation of state
parameter of the original one-component Chaplygin gas provides us
with modifications of the adiabatic sound speed. Furthermore, we
investigate the large scale perturbation dynamics and obtain a
non-adiabatic contribution to the gravitational potential which is
relevant for the ISW. The results of the paper are summarized in
Section \ref{conclusions}.

\section{Interacting two-fluid dynamics}
\label{two-fluid dynamics}

Much of the appeal of a Chaplygin gas model of the cosmic
substratum is due to the circumstance that it represents a unified
description of dark energy and dark matter within a one component
model. In order to describe non-adiabatic features which are the
manifestation of an internal structure of the medium it will be
instructive, however, to start the discussion by establishing
basic relations of a two-fluid description. The point here is that
we shall reveal an equivalent two-component picture of any
one-component Chaplygin gas (for different two-component
decompositions see \cite{bento2,koivisto}). This circumstance can
be used to introduce a simple model for an internal structure
which gives rise to non-adiabatic pressure perturbations.

\subsection{General setting}
\label{general}

We assume the cosmic medium to behave as a perfect fluid with an
energy-momentum tensor
\begin{equation}\label{Ttot}
T^{ik} = \rho u^{i}u^{k} + p h^{ik} ,\qquad\ h^{ik} = g^{ik} +
u^{i} u^{k}\ .
\end{equation}
Our approach is based on the decomposition
\begin{equation}\label{Tsum}
T^{ik} = T_{1}^{ik} + T_{2}^{ik}
\end{equation}
of the total energy-momentum into two parts with ($A= 1, 2$)
\begin{equation}\label{TA}
T_{A}^{ik} = \rho_{A} u_A^{i} u^{k}_{A} + p_{A}^{*} h_{A}^{ik} \
,\qquad\ h_{A}^{ik} = g^{ik} + u_A^{i} u^{k}_{A} , \qquad
T_{A\,;k}^{ik} = 0\ .
\end{equation}
For our purpose it is convenient to split the (effective)
pressures $p_{1}^{*}$ and $p_{2}^{*}$ according to
\begin{equation}\label{defp*}
p_{1}^{*} \equiv p_{1} + \Pi_{1}\ ,\quad p_{2}^{*} \equiv p_{2} +
\Pi_{2}\ .
\end{equation}
This procedure allows us to introduce an arbitrary coupling
between both fluids by the requirement
\begin{equation}\label{Pi1=-Pi2}
\Pi_{1} = -\Pi_2 \equiv \Pi \quad \Leftrightarrow \quad p_{1}^{*}
= p_{1} + \Pi\ ,\quad p_{2}^{*} = p_{2} - \Pi\ .
\end{equation}
Under this condition the apparently  separate energy-momentum
conservation $T_{A\,;k}^{ik} = 0$ in (\ref{TA}) is only formal.
Then, for a homogeneous and isotropic universe with $u_1^{i}  =
u^{i}_{2} = u^{i}$ and $u^{a}_{;a} = 3H$, where $H$ is the Hubble
rate, the individual energy balance equations are coupled and take
the form
\begin{equation}
\dot \rho_1 + 3H\left(\rho_1 + p_1 \right) = - 3H\Pi \quad
\Leftrightarrow \quad \dot \rho_1 + 3 H \left(\rho_1 +
p_1^{*}\right)= 0 \ , \label{dotrho1}
\end{equation}
and
\begin{equation}
\dot \rho_2 + 3 H \left(\rho_2 + p_2 \right) = 3H\Pi \quad
\Leftrightarrow \quad \dot \rho_2 + 3 H \left(\rho_2 +
p_2^{*}\right)= 0 \ . \label{dotrho2}
\end{equation}
Adding up these balances, consistently reproduces the total energy
conservation
\begin{equation}
\dot \rho+ 3 H \left(\rho+ p\right) = 0 \  \label{dotrho}
\end{equation}
with
\begin{equation}
\rho = \rho_1 + \rho_2\ ,\quad p= p_1 + p_2 = p_1^{*} + p_2^{*}\ .
\label{}
\end{equation}
 The total adiabatic sound velocity may be split according to
\begin{equation}
\frac{\dot{p}}{\dot{\rho}}  =
\frac{\dot{\rho}_{1}}{\dot{\rho}}\frac{\dot p _1}{\dot \rho _1} +
\frac{\dot{\rho}_{2}}{\dot{\rho}} \frac{\dot p _2}{\dot \rho _2} =
\frac{\dot{\rho}_{1}}{\dot{\rho}}\frac{\dot p^{*}_1}{\dot \rho _1}
+ \frac{\dot{\rho}_{2}}{\dot{\rho}} \frac{\dot p^{*}_2}{\dot \rho
_2} \ . \label{soundspeedsplit}
\end{equation}
We emphasize that so far the equations of state and the nature of
the interaction between the components are left unspecified.

\subsection{Matter perturbations}
\label{matter perturbations}

To study perturbations about the homogeneous and isotropic
background we introduce the following quantities. We define the
total fractional energy density perturbation
\begin{equation}
D \equiv  \frac{\hat{\rho}}{\rho+ p} = - 3H
\frac{\hat{\rho}}{\dot\rho}\ , \label{D}
\end{equation}
and  the energy density perturbations for the components ($A=1,\,
2$)
\begin{equation}
D_{A} \equiv \frac{\hat{\rho}_{A}}{\rho _{A}+ p_{A}^{*}} = - 3H
\frac{\hat{\rho}_A}{\dot\rho_A} \ .\label{DA}
\end{equation}
Likewise, the corresponding pressure perturbations are
\begin{equation}
P \equiv  \frac{\hat{p}}{\rho+ p} = - 3H \frac{\hat{p}}{\dot\rho}
= - 3H \frac{\dot{p}}{\dot\rho}\frac{\hat{p}}{\dot p}\ ,
\label{Ptot}
\end{equation}
and
\begin{equation}
P_{A} \equiv \frac{\hat{p}_{A}}{\rho _{A}+ p_{A}^{*}} = - 3H
\frac{\hat{p}_A}{\dot\rho_A} = - 3H
\frac{\dot{p}_{A}}{\dot{\rho}_A}\frac{\hat{p}_A}{\dot p_A}\ .
\label{PA}
\end{equation}
One also realizes that
\begin{equation}
D = \frac{\dot\rho_{1}}{\dot\rho} D_{1}+
\frac{\dot\rho_{2}}{\dot\rho}D_{2} \ \label{Dsum}
\end{equation}
and
\begin{equation}
P = \frac{\dot\rho_{1}}{\dot\rho} P_{1} +
\frac{\dot\rho_{2}}{\dot\rho} P_{2}  \ .\label{Psum}
\end{equation}
In terms of these quantities the non-adiabatic pressure
perturbations
\begin{equation}
P - \frac{\dot{p}}{\dot{\rho}}D = - 3H\frac{\dot{p}}{\dot\rho}
\left[\frac{\hat{p}}{\dot p} -  \frac{\hat{\rho}}{\dot\rho}\right]
\ \label{defPnad}
\end{equation}
are then generally characterized by
\begin{eqnarray}
P - \frac{\dot{p}}{\dot{\rho}}D &=& \frac{\dot\rho_{1}}{\dot\rho}
\left(P_1 - \frac{\dot{p}_1}{\dot{\rho}_1}D_1 \right) +
\frac{\dot\rho_{2}}{\dot\rho}
\left(P_2 - \frac{\dot{p}_2}{\dot{\rho}_2}D_2 \right)\nonumber\\
&& + \frac{\dot\rho_{1}\dot\rho_{2}}{\left(\dot\rho \right)^2}
\left[\frac{\dot{p}_2}{\dot{\rho}_2} -
\frac{\dot{p}_1}{\dot{\rho}_1} \right] \left[D_2 - D_1\right]\ .
\label{Pnad}
\end{eqnarray}
\ \\
The first two terms on the right-hand side describe internal
non-adiabatic perturbations within the individual components. The
last term takes into account non-adiabatic perturbations due to
the two-component nature of the medium.

\subsection{Metric perturbations}

A homogenous and isotropic background universe with scalar metric
perturbations is characterized by the line element
\begin{equation}
\mbox{d}s^{2} = - \left(1 + 2 \phi\right)\mbox{d}t^2 + 2 a^2
F_{,\alpha }\mbox{d}t\mbox{d}x^{\alpha} +
a^2\left[\left(1-2\psi\right)\delta _{\alpha \beta} + 2E_{,\alpha
\beta} \right] \mbox{d}x^\alpha\mbox{d}x^\beta \ .\label{metric}
\end{equation}
The components of the perturbed 4-velocity are
\begin{equation}
\hat{u}^0 = \hat{u}_0  = - \phi \label{u0}
\end{equation}
and
\begin{equation}
a^2\hat{u}^\mu + a^2F_{,\mu} = \hat{u}_\mu \equiv v_{,\mu}\ .
\label{umu}
\end{equation}
Furthermore, we define
\begin{equation}
\chi \equiv a^2\left(\dot{E} -F\right) \ .\label{chi}
\end{equation}
The perturbation dynamics is most conveniently described in terms
of the gauge-invariant variable \cite{Bardeenetal83}
\begin{equation}
\zeta \equiv -\psi + \frac{1}{3}\frac{\hat{\rho}}{\rho + p} =
-\psi - H \frac{\hat{\rho}}{\dot\rho}\ .
 \label{defzeta}
\end{equation}
Corresponding quantities for the components are
\begin{equation}
\zeta_A \equiv -\psi + \frac{1}{3}\frac{\hat{\rho}_A}{\rho_A +
p^{*}_A} = - \psi - H \frac{\hat{\rho}_A}{\dot\rho _A}\ .
 \label{defzetaA}
\end{equation}
The variable $\zeta$ obeys the equation (cf
\cite{Lyth,GarciaWands,Wandsetal00})
\begin{equation}
\dot\zeta = - H \left(P - \frac{\dot{p}}{\dot{\rho}}D\right) -
\frac{1}{3a^2}\Delta v^{\chi} \ ,
 \label{dotzetageneral}
\end{equation}
where
\begin{equation}
v^{\chi} \equiv v + \chi
 \label{vchi}
\end{equation}
is the gauge-invariant velocity potential. Using in
(\ref{dotzetageneral}) the formula (\ref{Pnad}) for $P-(\dot
p/\dot \rho)D$ together with the definitions (\ref{defzetaA}) and
taking into account that $p_1 = p_1\left(\rho_1\right)$ and $p_2 =
p_2\left(\rho_2\right)$, Eq.~(\ref{dotzetageneral}) reduces to
\begin{equation}
\dot\zeta =  - 3H \frac{\dot\rho _1 \dot\rho _2}{\dot\rho ^2 }
\left[ \frac{\dot p _2}{\dot\rho _2} -  \frac{\dot  p_1}{\dot\rho
_1} \right] \left( \zeta_2 - \zeta_1 \right) -
\frac{1}{3a^2}\Delta v^{\chi}\ . \label{dotzeta}
\end{equation}
Equation (\ref{dotzeta}) expresses the general fact that entropy
perturbations $\left( \zeta_2 - \zeta_1 \right)$ source curvature
perturbations $\zeta$. This completes our preliminary definitions
and remarks which, under the given conditions, are generally
valid. No specific Chaplygin gas properties have been used so far.

\section{The Chaplygin gas}
\label{Chaplygin}

\subsection{Two-component interpretation}
\label{two-component}

The Chaplygin gas is characterized by an equation of state \be p =
- \frac{A}{\rho}\ , \label{eosChap} \ee where $p$ is the pressure
and $\rho$ is the energy density of the gas. $A$ is a positive
constant. This equation of state gives rise to the energy density
\cite{Kamen} \be \rho = \sqrt{A +  \frac{B}{a^6}}\ ,
\label{rhoChap} \ee where $B$ is another (positive) constant. Now,
let us rename the constants according to \be A =
A_2\left(1+\kappa\right)\ , \qquad B = B_2\left(1+\kappa\right)^2\
. \label{defA2} \ee For any constant, non-negative $\kappa$ the
quantities $A_2$ and $B_2$ are constant and non-negative as well.
So far, no physical meaning is associated with these constants. It
is obvious, that this split allows us to write the energy density
(\ref{rhoChap}) as \be \rho =
\left(1+\kappa\right)\sqrt{\frac{A_2}{1+ \kappa} +
\frac{B_2}{a^6}}\ . \label{rhoviaA2} \ee This means, $\rho$ can be
regarded as consisting of two components, $\rho_{1}$ and
$\rho_{2}$, \be \rho = \rho_1 + \rho_2\ , \label{5} \ee with \be
\rho_1 = \sqrt{\frac{A_2\kappa^2}{1+ \kappa} +
\frac{B_2\kappa^2}{a^6}}\ \label{rho1} \ee and \be \rho_2 =
\sqrt{\frac{A_2}{1+ \kappa} +  \frac{B_2}{a^6}} \ , \label{rho2}
\ee where $\rho_1 = \kappa\rho_2$ is valid, i.e., the ratio
$\rho_1/\rho_2 = \kappa$ is constant. Obviously, both these
expressions (\ref{rho1}) and (\ref{rho2}) describe energy
densities for Chaplygin gases again. The corresponding (effective)
equations of state are (the notations are chosen in agreement with
the formalism of subsection \ref{general}) \be p_1^{*} = -
\frac{A_2 \kappa^2}{\left(1+\kappa\right)\rho_1} = - \frac{A_2
\kappa}{\left(1+\kappa\right)\rho_2} \ , \label{p1star} \ee and
\be p_2^{*} =  - \frac{A_2}{\left(1+\kappa\right)\rho_2} \ .
\label{p2star} \ee This implies $p_1^{*} = \kappa p_2^{*}$, i.e.,
the effective pressures of both components differ by the same
constant which also characterizes the ratio of both energy
densities. It is straightforward to check that \be p_1^{*} +
p_2^{*} = - \frac{A_2}{\rho_2} = - \frac{A}{\rho} = p \ .
\label{psum} \ee It is further convenient to introduce a quantity
\begin{equation}
p_2 \equiv - \frac{A_2}{\rho_2}\ . \label{p2}
\end{equation}
This quantity differs from $p_2^{*}$ by
\begin{equation}
p_2^{*} - p_2 = \frac{\kappa A_2}{\left(1+\kappa\right)\rho_2} = -
p_1^{*}\ . \label{12}
\end{equation}
Defining also a quantity $\Pi$ by
\begin{equation}
\Pi \equiv p_1^{*} = - \frac{\kappa}{1+\kappa}\frac{A_2}{\rho_2} =
- \frac{\kappa}{1+\kappa}\frac{A}{\rho}\ , \label{Pip1star}
\end{equation}
one checks by direct calculation that the following relations are
valid:
\begin{equation}
\dot \rho_1 + 3 H \rho_1 = - 3H\Pi \quad \Leftrightarrow \quad
\dot \rho_1 + 3 H \left(\rho_1 + p_1^{*}\right)= 0 \ ,
\label{dotrho1n}
\end{equation}
and
\begin{equation}
\dot \rho_2 + 3 H \left(\rho_2 + p_2 \right) = 3H\Pi \quad
\Leftrightarrow \quad \dot \rho_2 + 3 H \left(\rho_2 +
p_2^{*}\right)= 0 \ . \label{dotrho2n}
\end{equation}
Adding up these balances, consistently reproduces the total energy
conservation
\begin{equation}
\dot \rho+ 3 H \left(\rho+ p\right) = 0 \ . \label{dotrhon}
\end{equation}
At this point it becomes clear what we have obtained by the formal
manipulations in Eq.(\ref{defA2}). The set of equations
(\ref{dotrho1n}) - (\ref{dotrhon}) fits exactly the two-component
description (\ref{dotrho1}) - (\ref{dotrho}). It describes a
two-component system in which the components interact with each
other. The role of the interaction is to keep the ratio $\kappa$
of the energy densities of both components constant. In other
words, {\it any given Chaplygin gas can be thought as being
composed of another Chaplygin gas which is in interaction with a
pressureless fluid such that the energy density ratio of both
components is fixed.} Apparently, this admits a further
decomposition of the resulting Chaplygin gas component $2$ in
which now $A_2$ plays the role that $A$ played so far, into still
another Chaplygin gas interacting with matter in a similar way
etc. It will turn out, however, that this formal possibility is
not relevant for our approach.

\subsection{Chaplygin gas interacting with matter}

For our further considerations it is essential that the entire
reasoning of the previous subsection \ref{two-component} may be
reversed. In such a case the starting point is a mixture of a
Chaplygin gas and a pressureless fluid which are in interaction
with each other according to
\begin{equation}
\dot \rho_1 + 3 H \rho_1 = - 3H\tilde{\Pi} \ , \label{dotrho13}
\end{equation}
\begin{equation}
\dot \rho_2 + 3 H \left(\rho_2 + p_2 \right) = 3H\tilde{\Pi} \ ,
\label{dotrho23}
\end{equation}
with
\begin{equation}
p_{2} = - \frac{A_{2}}{\rho_{2}}\ =\ p \label{p23}
\end{equation}
and an (initially unspecified) interaction quantity $\tilde{\Pi}$.
This is a special case of (\ref{dotrho1}), (\ref{dotrho2}). For
the time dependence of the (at this moment still arbitrary) ratio
$\kappa \equiv \rho_1/\rho_2$ we have
\begin{equation}
\dot \kappa  = \kappa \left[\frac{\dot \rho _1}{\rho_1} -
\frac{\dot \rho _2}{\rho_2}\right] = 3H\kappa
\left[\frac{p_2}{\rho_2} -
\frac{\rho}{\rho_1\rho_2}\tilde{\Pi}\right] \ . \label{dotkappa}
\end{equation}
The stationarity requirement for $\kappa$ then implies
\begin{equation}
\dot \kappa  = 0 \quad\Leftrightarrow \quad \tilde{\Pi} =
\frac{\rho_1}{\rho}p_2 \ . \label{dotkappa0}
\end{equation}
With (\ref{p23}) and $\rho_1/\rho = \kappa/(1 + \kappa)$ one
recovers relation (\ref{Pip1star}) if $\tilde{\Pi}$ is identified
with $\Pi$, i.e.,
\begin{equation}
\tilde{\Pi} = \Pi = -
\frac{\kappa}{1+\kappa}\frac{A_{2}}{\rho_{2}} = -
\frac{\kappa}{1+\kappa}\frac{A}{\rho} = -
\kappa\frac{A_{2}}{\rho}\ . \label{PiviaA2}
\end{equation}
Consequently, the energy densities of the components are again
given by (\ref{rho1}) and (\ref{rho2}) and add up to
(\ref{rhoviaA2}) which, by redefining the constants according to
(\ref{defA2}), results in the energy density (\ref{rhoChap}).

To summarize, we may either follow the line of the previous
subsection \ref{two-component} and find that one may attribute a
specific two-component interpretation to any given Chaplygin gas
or we may, alternatively, assert that a system consisting of a
Chaplygin gas and a pressureless fluid which are in interaction
with each other according to (\ref{dotrho13}), (\ref{dotrho23})
and (\ref{dotkappa0}), is equivalent to another Chaplygin gas with
the energy density (\ref{rhoChap}).

The components have sound velocities which are different from each
other and different from the overall sound velocity. For the
latter we have
\begin{equation}
\frac{\dot p}{\dot \rho} = - \frac{p}{\rho} = \frac{A}{\rho^2} =
\frac{A}{A + \frac{B}{a^6}} \ . \label{soundChap}
\end{equation}
The sound velocities of the components are $\dot p _1 /\dot \rho
_1 = 0$ and
\begin{equation}
\frac{\dot p _2}{\dot \rho _2} = - \frac{p_2}{\rho_2} =
\frac{A_2}{\rho_2^2} = \frac{A_2}{\frac{A_2}{1 + \kappa} +
\frac{B_2}{a^6}} = \left(1  + \kappa\right)\frac{\dot p}{\dot
\rho} \ . \label{sound2}
\end{equation}
Since for large times we have $\dot{p}/\dot{\rho} \rightarrow 1$,
the quantity $\dot{p}_{2}/\dot{\rho}_2$ may be larger than unity
which at the first glance seems to imply a superluminal sound
propagation. However, $\dot{p}_{2}/\dot{\rho}_2$ is a formal
quantity only, which does not describe any propagation phenomenon.
The adiabatic sound speed (\ref{soundChap}) may also be split with
respect to the effective presures according to
(\ref{soundspeedsplit}) with
\begin{equation}
\frac{\dot p^{*}_1}{\dot \rho _1} =
\frac{\dot{p}}{\dot{\rho}}\qquad {\rm and} \qquad \frac{\dot
p^{*}_2}{\dot \rho _2} = \frac{\dot{p}}{\dot{\rho}}  \ .
\label{23c}
\end{equation}
These effective sound velocities of the components coincide with
the total sound velocity.

\section{Non-adiabatic pressure perturbations}
\label{non-adiabatic}

\subsection{Perturbation dynamics}
\subsubsection{Effective sound speed}
\ \\
The Chaplygin in its two-component interpretation of the
previous section belongs to a class of models with
\begin{equation}
\dot{\kappa} = 0 \quad {\rm and}\quad \rho_1 + p_1^{*} = \kappa
\left(\rho_2 + p_2^{*}\right)\ . \label{constkappa}
\end{equation}
Generally, the components may have  (not necessarily constant)
equations of state
\\
\begin{equation}
p_{1} = w_{1} \rho_{1}\quad{\rm and}\quad p_{2} = w_{2} \rho_{2}\
. \label{}
\end{equation}
For the total equation of state parameter $w$ it follows that
\begin{equation}
w =  \frac{w_2 + \kappa w_1}{1 + \kappa}\quad{\rm and}\quad p = w
\rho \ . \label{}
\end{equation}
For our special Chaplygin gas case these quantities  are
\begin{equation}
 w_1 = 0 ,\quad w_2 = - \frac{A_2}{\rho_{2}^{2}} ,\quad
 w = - \frac{A}{\rho^{2}}
\ . \label{w}
\end{equation}
Under the condition (\ref{constkappa}) the difference of the
individual sound speeds is
\begin{equation}
\frac{\dot{p}_2}{\dot{\rho}_2} - \frac{\dot{p}_1}{\dot{\rho}_1} =
\frac{1 + \kappa}{\kappa} \frac{\dot{\Pi}}{\dot{\rho}_2} \ .
\label{}
\end{equation}
Introducing now matter perturbations in terms of the quantities
defined in subsection \ref{matter perturbations} and allowing the
density ratio $\kappa$ to fluctuate, the difference between the
fractional energy density perturbations which describes entropy
perturbations becomes directly proportional to $\hat \kappa$:
\begin{equation}
D_2 - D_1 = 3\left(\zeta_2 - \zeta_1\right) = -  \frac{1}{1 +
w}\,\frac{\hat \kappa}{\kappa} \ . \label{D2-D1}
\end{equation}
This demonstrates, that a perturbation of the density ratio is
essential for entropy perturbations to occur. Of course, for a
\anf true" constant $\kappa$ there are no non-adiabatic
contributions  and the two-component description does not give
anything new compared to the original one-component picture. It
should be mentioned that, since $\dot \kappa = 0$, the quantity
$\hat \kappa$ is gauge-invariant (as is the combination on the
left-hand side). Since the background ratio $\kappa$ also
determines the relation between the equation of state parameters
$A$ and $A_2$ in (\ref{defA2}) a further specification may be
performed. We shall assume from now on the parameter $A_2$ to be a
true constant, i.e., $\hat{A}_2 =0$. This choice implies
\begin{equation}
\frac{\hat{A}}{A} = \frac{\hat{\kappa}}{1 + \kappa} \ .
\label{hatAChap}
\end{equation}
Within this model a fluctuation of the density ratio is equivalent
to a fluctuation of the equation of state parameter $A$. Any
fluctuation of $\kappa$ or $A$  generates a non-adiabatic pressure
fluctuation
\begin{equation}
P - \frac{\dot{p}}{\dot{\rho}}D =
\frac{w}{1+w}\,\frac{\hat{\kappa}}{1 + \kappa} =
\frac{w}{1+w}\,\frac{\hat{A}}{A}\ . \label{PnaChap}
\end{equation}
We conclude that perturbing the equation of state parameter $A$
represents a way to introduce non-adiabatic pressure perturbations
in a Chaplygin gas. It should be mentioned that this type of
perturbation can be obtained in a much simpler way by admitting
fluctuations of $A$ in the equation of state (\ref{eosChap})
itself. It is the most obvious choice which strictly speaking does
not depend on the assumption of a specific internal structure.
Nevertheless, we believe the two fluid context chosen here to be
superior since it is connected with a model for the origin of the
fluctuations of the equation of state parameter. Moreover, it is
expedient to recall that in an underlying string theoretical
formalism the parameter $A$ is connected with the interaction
strength of d-branes \cite{Jackiw}. Hence, a fluctuating $A$
corresponds to fluctuating interactions in string theory. An
alternative, perhaps more transparent way of understanding the
meaning of the fluctuation of the parameter $A$ is to consider the
Born-Infeld action that leads to the Chaplygin gas. The Lagrangian
density in this case takes the form \cite{sen,copeland}
\begin{equation}
L = \sqrt{-g}V(\phi)\sqrt{- det[g_{\mu\nu} +
\phi_{;\mu}\phi_{;\nu}]} \quad , \label{LBI}
\end{equation}
where $V(\phi)$ is a potential term. This Lagrangian density leads
to the Chaplygin gas equation of state for \cite{bento1,bento2}
\begin{equation}
V(\phi) = \sqrt{A}\ . \label{VA}
\end{equation}
For $\phi = \phi_0 + \hat{\phi}$, the fluctuation in $A$ is given
by
\begin{equation}
\frac{\hat{A}}{A}= 2 \left(\frac{V'}{V}\right)_{\phi =
\phi_0}\hat{\phi} \quad ,
\end{equation}
where the prime denotes the derivative with respect to $\phi$.
Consequently, fluctuations of $A$ are allowed if the scalar field
$\phi$ is not in the minimum of the potential. It has been shown
that there are configurations for which a sufficiently flat
potential (equivalent to an almost constant $A$) admits
accelerated expansion \cite{copeland}.

For vanishing fluctuations of $\kappa$ and $A$ in (\ref{PnaChap})
we recover the adiabatic case. Given that $w<0$, any $\hat{A}
> 0$ will reduce the adiabatic pressure perturbations.
 For definite statements further assumptions about
$\hat{A}$ (or $\hat{\kappa}$) are necessary since otherwise the
problem is undetermined. A simple case which will be used below is
$\hat{A}/A=$ constant. The structure of (\ref{PnaChap}) also
motivates a choice $\hat{A}/A = \mu\left(1 + w\right) D$, for
which
\begin{equation}
P = c_{eff}^{2} D \ ,\quad c_{eff}^{2} \equiv
\frac{\dot{p}}{\dot{\rho}}\left(1 - \mu\right)\ . \label{Pceff}
\end{equation}
Any $\mu$ in the range $0<\mu\leq 1$ leads to an effective sound
speed square $c_{eff}^{2}$ which is reduced compared to the
adiabatic value $ \dot{p}/\dot{\rho} = -w$. It is interesting to
note that the relation
\[
\hat{p} = \left(1 - \mu\right)\frac{\dot{p}}{\dot{\rho}}\hat{\rho}
= - \left(1 - \mu\right)\frac{p}{\rho}\hat{\rho} \ ,
\]
which is a  different way of writing (\ref{Pceff}), coincides with
that of a generalized Chaplygin gas with an equation of state $p =
- C/\rho^{\left(1 - \mu\right)}$ with (a true) constant $C$. For a
generalized Chaplygin gas one would, however, also have $\dot{p} =
- \left(1 - \mu\right) \frac{p}{\rho}\dot{\rho}$ and hence
adiabatic perturbations only. In a sense, our strategy to consider
perturbations $\hat{A}\neq 0$ while $\dot{A} =0$, implies that the
medium behaves as a Chaplygin gas in the background and shares
features of a generalized Chaplygin gas on the perturbative level.

At this point also the role of the assumption $\hat{A}_2 =0$ that
precedes eq.~(\ref{hatAChap}) becomes clear. We mentioned at the
end of subsection \ref{two-component} that the decomposition of a
given Chaplygin gas into another Chaplygin gas interacting with
matter may be repeated again and again. In a subsequent step $A_2$
would play the role that $A$ played so far. Hence, the assumption
$\hat{A}_2 =0$ ensures that no further non-adiabatic contributions
will appear.

As already mentioned, a different two-fluid model for a
(generalized) Chaplygin gas has been developed in reference
\cite{bento2}, where this gas has been decomposed into a
pressureless fluid and a time dependent cosmological ``constant"
with interactions between them. These authors obtained a
modification of the perturbation dynamics through the interaction
as well. The main difference to our approach is that in
\cite{bento2} the (generalized) Chaplygin gas was assumed to
decompose into two components with constant equation of state
parameters but variable ratio of the energy densities. Our
splitting neither assumes the existence of a pressureless
component from the outset nor does it restrict the equation of
state parameters to have constant values. The split (\ref{5}) -
(\ref{rho2}) which leads to (\ref{Pip1star}) - (\ref{dotrhon})
[with (\ref{eosChap}) and (\ref{rhoChap})] does not imply any a
priori requirements on the equation of state parameters of the
components. The equation of state parameters in (\ref{w}) are not
imposed but emerge in our splitting procedure. No phantomlike
contribution does occur in our setting which, moreover, is
characterized by a different sign of the energy flow between the
components. This shows that the claim of the authors of
\cite{bento2} to present a {\it unique} decomposition of the
(generalized) Chaplygin gas into DM and DE is not generally true.
It is only valid under the {\it additional} assumptions $p_{dm}=0$
and $w_X =$ const. Due to the circumstance that they deal with a
$\Lambda$ like fluid as DE component, perturbations of the latter
do not occur in their analysis which seems to exclude
perturbations of the parameter $A$.

\subsubsection{Fluctuating decay rate}
\ \\
There exists still another interpretation for the appearance
of non-adiabatic perturbations. The sign of the effective pressure
$\Pi$ in (\ref{PiviaA2}) indicates that within the two-component
interpretation of the Chaplygin gas there is a transfer of energy
from component 2 to component 1. This transfer may be described in
terms of a decay rate $\Gamma$, defined by
\begin{equation}
3H\Pi \equiv - \Gamma\rho_2  \  , \label{intrGamma}
\end{equation}
with the help of which the balance (\ref{dotrho23}) may be written
as
\begin{equation}
\dot{\rho}_{2} + 3H\left(1 + w_2\right)\rho_{2} = -
\Gamma\rho_{2}\ . \label{dotrho2Gamma}
\end{equation}
From (\ref{PiviaA2}) we obtain $\Pi \rho = - \kappa A_2
\Rightarrow \left(\Pi \rho\right)^{\displaystyle\cdot} = 0$. On
the other hand, (\ref{intrGamma}) implies $\Pi \rho = -
\left(\Gamma/3 H\right)\rho_2 \rho$, i.e., we have
\begin{equation}
\Pi \rho = - \frac{\Gamma}{3H} \frac{\rho^{2}}{1 + \kappa}\  .
\label{Pirho}
\end{equation}
Since in a spatially flat universe $\rho \propto H^{2}$ by
Friedmann's equation, it follows that
\begin{equation}
\left(\Pi \rho\right)^{\displaystyle\cdot} = 0 \quad
\Rightarrow\quad \left(\Gamma H^{3}\right)^{\displaystyle\cdot} =
0 \quad \Rightarrow\quad \Gamma\propto H^{-3}\ . \label{dotPirho}
\end{equation}
Consequently, a stationary ratio of the energy densities is
guaranteed for a decay rate which scales with the inverse third
power of the Hubble rate. {\it In the background, a Chaplygin gas
with equation of state $p =- A/\rho$ can always be modelled as a
mixture of a Chaplygin gas with equation of state $p_2 =-
A_2/\rho_2$ and matter with $p_1 \ll \rho_1$, in which the
Chaplygin gas component decays into the matter component at a rate
$\Gamma\propto H^{-3}$ .} Fluctuations of this decay rate are
accompanied by fluctuations $\hat{\kappa} $ of the energy density
ratio. Since $\hat{A}_2 =0$ was assumed, we have
\begin{equation}
\left(\Pi \rho\right)^{\hat{}} = - \hat{\kappa} A_2 \quad
\Rightarrow\quad \frac{\left(\Pi \rho\right)^{\hat{}}}{\Pi \rho} =
\frac{\hat{\kappa}}{\kappa}\ . \label{hatPirho}
\end{equation}
An alternative expression, derived from (\ref{Pirho}), is
\begin{equation}
\frac{\left(\Pi \rho\right)^{\hat{}}}{\Pi \rho} =
\left[\frac{\hat{\Gamma}}{\Gamma} - \frac{\hat{H}}{H} +
\frac{2\hat{\rho}}{\rho} - \frac{\hat{\kappa}}{1 + \kappa}\right]\
. \label{hatPirho1}
\end{equation}
(Here $\hat{H}$ is defined as $\hat{\Theta}/3$, where $\Theta
\equiv u^{a}_{;a}$ is the fluid expansion). On large perturbation
scales for which spatial gradients may be neglected, the first
order $G^{0}_{0}$ field equation implies
\begin{equation}
\frac{\hat{\rho}}{\rho} = 2\frac{\hat{H}}{H} \ ,
\label{hatrhohatT}
\end{equation}
i.e.,
\begin{equation}
\frac{\left(\Pi \rho\right)^{\hat{}}}{\Pi \rho} =
\frac{\left(\Gamma H^3\right)^{\hat{}}}{\Gamma H^3}  -
\frac{\hat{\kappa}}{1 + \kappa}\ . \label{hatPirho2}
\end{equation}
Combination of the two expressions (\ref{hatPirho}) and
(\ref{hatPirho2}) for $\left(\Pi \rho\right)^{\hat{}}/\left(\Pi
\rho\right)$ yields
\begin{equation}
\frac{\left(\Gamma H^3\right)^{\hat{}}}{\Gamma H^3}  = \frac{2 +
\kappa}{1 + \kappa}\,\frac{\hat{\kappa}}{\kappa}\ . \label{hatGT}
\end{equation}
Large scale fluctuations about $\Gamma \propto H^{-3}$ correspond
to fluctuations of $\kappa$ which, in turn, are equivalent to
fluctuations of $A$ according to (\ref{hatAChap}).

\subsubsection{Dynamics on large perturbation scales}
\ \\
Using the expression (\ref{PnaChap}) in (\ref{dotzetageneral})
we may now consider the large scale perturbation dynamics. If we
neglect spatial gradients, the equation for $\zeta$ reduces to
\begin{equation}
\dot{\zeta} = - H \frac{w}{1 + w}\frac{\hat{\kappa}}{1 + \kappa} =
-  H \frac{w}{1 + w}\frac{\hat{A}}{A}\ . \label{dotzetalarge}
\end{equation}
For a qualitative discussion of the impact of a non-vanishing
fluctuation of the parameter $A$ on the perturbation dynamics we
assume $\hat{A}/A =$ constant as the simplest choice. With the
Chaplygin gas equation of state
\begin{equation}
w = - \frac{Aa^{6}}{Aa^{6} + B} \qquad \Rightarrow \qquad 1 + w =
\frac{B}{Aa^{6} + B}\label{w,1+w}
\end{equation}
we obtain
\begin{equation}
\zeta = \zeta_{i} +
\frac{1}{6}\frac{Aa^{6}}{B}\frac{\hat{A}}{A}\,. \label{zeta=}
\end{equation}
The adiabatic case $\zeta = \zeta_{i}$ is, of course recovered for
$\hat{A} = 0$. For $B \gg Aa^{6}$, i.e., for small $a$ with a
negligible pressure we have $\zeta \approx \zeta_{i}$. In the
opposite limit $Aa^{6} \gg B$ the non-adiabatic contribution can
be substantial and may dominate the entire dynamics.

\subsection{The gravitational potential}

A time dependent equation of state and/or a varying $\zeta$ due to
non-adiabatic perturbations generate a time dependence in the
gravitational potential which, in turn, gives rise to the ISW.
With the gauge-invariant definitions
\begin{equation}
\psi^{\chi} \equiv \psi + H\chi\ ,\quad \psi^{v} \equiv \psi - H
v\ ,\quad \phi^{\chi} \equiv \phi - \dot{\chi}\ ,\label{defpsichi}
\end{equation}
where $\psi^{v}$ is the comoving curvature perturbation, the
relations (cf (4.92) in \cite{LL})
\begin{equation}
\zeta + \psi^{v} = \frac{2}{9\left(1 + w\right)} \frac{\Delta
\psi^{\chi}}{H^2 a^2}\ ,\quad \phi^{\chi} =
\psi^{\chi}\label{zetapsi}
\end{equation}
and (cf (4.169) in \cite{LL})
\begin{equation}
\dot{\psi}^{\chi} + \frac{3}{2}H\left(\frac{5}{3} +
w\right)\psi^{\chi} = \frac{3}{2}\left(1 + w\right) H \psi^{v} \
,\label{dotpsi}
\end{equation}
are generally valid. (The upper indices $\chi$ and $v$ are kept to
indicate the gauge-invariant character of the introduced
quantities). From the definitions (\ref{defzeta}) for $\zeta$ and
(\ref{defpsichi}) for $\psi^{v}$ it is obvious, that the
combination $\zeta + \psi^{v}$ on the left-hand side of the first
relation (\ref{zetapsi}) describes comoving, fractional energy
density perturbations.  On large perturbation scales we may
neglect the gradient term in (\ref{zetapsi}) and $\psi^{v}$ in
(\ref{dotpsi}) may be replaced by $\zeta$:
\begin{equation}
\psi^{v} \approx - \zeta \quad\Rightarrow \quad \dot{\psi}^{\chi}
+ \frac{3}{2}H\left(\frac{5}{3} + w\right)\psi^{\chi} = -
\frac{3}{2}\left(1 + w\right) H \zeta\ .\label{dotpsi1}
\end{equation}
A non-vanishing time dependence of the gravitational potential
$\psi^{\chi}$ generates the ISW. The impact of the dark energy
sound speed on the ISW has recently been  discussed for
quintessential models \cite{Erickson,DeDeo,Bean2}. For
(\ref{w,1+w}) and (\ref{zeta=}) the equation for $\psi^{\chi}$ in
~(\ref{dotpsi1}) has the solution
\begin{eqnarray}
\psi^{\chi} = - \frac{3}{2}\zeta_{i}\frac{B}{a}E^{-1/4}\int
\frac{\mbox{d}a}{Aa^{6}+B}E^{1/4}\qquad\qquad\qquad&& \nonumber\\-
\frac{1}{4} \frac{\hat{A}}{A}\frac{1}{a}E^{-1/4}\int
\mbox{d}a\frac{Aa^{6}}{Aa^{6}+B}E^{1/4} ,\label{solpsi}
\end{eqnarray}
where
\begin{equation}
E \equiv
\frac{Aa_{i}^{6}+B}{Aa^{6}+B}\left(\frac{a}{a_{i}}\right)^{6}
.\label{}
\end{equation}
The subscript $i$ denotes some reference epoch with $a_{i} \ll a$
in the era $B \gg Aa_{i}^{6} $. The first term in (\ref{solpsi})
is an adiabatic contribution. The fluctuations in $A$ give rise to
a second, non-adiabatic gravitational potential term. For a
qualitative discussion it is useful to consider the limit of very
large values of the scale factor, i.e., for $Aa^{6} \gg B$, where
the background equation of state is close to that corresponding to
a cosmological constant. In this realm the solution (\ref{solpsi})
is approximately
\begin{equation}
\psi^{\chi} \approx -
\frac{3}{10}\zeta_{i}\frac{a_{i}}{a}\frac{B}{Aa_{i}^{6}} -
\frac{1}{4} \frac{\hat{A}}{A} .\label{psilarge}
\end{equation}
The first term decays with $a^{-1}$ as a result of the time
dependence of the equation of state. However, due to the
non-adiabatic pressure perturbation there exists a constant second
term which sets a threshold below which $\psi^{\chi}$ does not
further decay. For sufficiently large values of the scale factor
the first term becomes negligible and the gravitational potential
approaches a constant value.

It is instructive to compare the potential (\ref{psilarge}) with
the corresponding result of the $\Lambda$CDM model. With
\begin{equation}
\rho_{\Lambda CDM} = \rho_{M} + \rho_{\Lambda} , \qquad \rho_{M}
=\rho_{M_i}\left(\frac{a_{i}}{a}\right)^{3} , \quad \rho_{\Lambda}
= {\rm const} , \label{}
\end{equation}
where the subindex $i$ now denotes an initial value with
$\rho_{M_i} \gg \rho_{\Lambda}$ and
\begin{equation}
 p_M \ll \rho_{M} ,\quad p_{\Lambda} = -
\rho_{\Lambda} , \quad \Rightarrow \quad w_{\Lambda CDM} = -
\frac{1}{1 +
\frac{\rho_{M_i}}{\rho_{\Lambda}}\left(\frac{a_{i}}{a}\right)^{3}}\
, \label{}
\end{equation}
we have $\zeta = \zeta_{i}$, i.e., $\zeta$ is a conserved
quantity. The gravitational potential in the long time limit $a
\gg a_{i}$ becomes
\begin{equation}
\psi_{\Lambda CDM}^{\chi} \approx -
\frac{3}{4}\zeta_{i}\frac{\rho_{M_i}}{\rho_{\Lambda}}
\frac{a_{i}}{a} .\label{psilcdm}
\end{equation}
In this case there does not exist a threshold value,
$\psi_{\Lambda CDM}^{\chi}$ simply decays. Since the change in the
gravitational potential affects the anisotropy spectrum at the
largest scales one expects a discrimination between models with
different values of $\hat{A}/A$.

\section{Conclusions}
\label{conclusions}

We have demonstrated that any Chaplygin gas in a homogeneous and
isotropic Universe can be regarded as being composed of another
Chaplygin gas which is in interaction with a pressureless fluid
such that that the background ratio of the energy densities of
both components remains constant. This two-component
interpretation allowed us to establish a simple model for
non-adiabatic pressure perturbations as the result of a specific
internal structure of the substratum. Within the one-component
picture this internal structure manifests itself as a spatial
fluctuation of the parameter $A$ in the Chaplygin gas equation of
state $p = - A/\rho$. Such fluctuations, which can be traced back
to fluctuations of a tachyon field potential, modify the adiabatic
sound speed of the medium which may shed new light on the status
of Chaplygin gas models of the cosmic substratum. For the special
case of a constant fractional perturbation $\hat{A}/A$ on large
scales we calculated the deviation of the quantity $\zeta$, which
describes curvature perturbations on constant density
hypersurfaces, from its (constant) adiabatic value. Furthermore,
fluctuations $\hat{A}$ give rise to an additional contribution to
the gravitational potential which we have determined. In the long
time limit when the Chaplygin gas equation of state is close to
that corresponding to a cosmological constant, this contribution
sets a threshold value below which the gravitational potential
does not decay. This is different from both the adiabatic
Chaplygin gas and the $\Lambda$CDM model with purely decaying
potentials and should be a potential tool to discriminate between
different dark energy models via different ISW contributions to
the lowest multipoles in the anisotropy spectrum of the cosmic
microwave background. A more quantitative analysis of this feature
will be given in a subsequent paper.\\
\ \\
{\bf Acknowledgements}\\
This work was partially supported by the Deutsche
Forschungsgemeinschaft. Discussions with J\'er\^ome Martin and
Nelson Pinto-Neto are gratefully acknowledged. JCF thanks the
Institut f\"ur Theoretische Physik, Universit\"at zu K\"oln for
hospitality during part of this work. WZ is grateful for
hospitality by the Departamento de F\'{\i}sica Universidade
Federal do Esp\'{\i}rito Santo.\\
\ \\
\ \\

\end{document}